\documentclass{article}
\usepackage{waspaa17,amsmath,graphicx,url,times}
\usepackage{color}
\usepackage{psfrag,amsmath,multirow,array,bm,cite,setspace}
\usepackage[colorlinks=true,linkcolor=black,anchorcolor=black,citecolor=black,urlcolor=cyan,bookmarks=false]{hyperref}

\newcommand{\stz}{\rule{0mm}{2.3ex}}

\newcommand{\figref}[1]{\figurename\,\ref{#1}}

\newcommand{\tabref}[1]{Table~\ref{#1}}

\definecolor{myblue}{rgb}{0, 0, 1}

\title{A Perceptual Weighting Filter Loss \\ for DNN Training in Speech Enhancement}

\name{Ziyue Zhao, Samy Elshamy, Tim Fingscheidt\thanks{The author Ziyue Zhao would like to thank China Scholarship Council (CSC) for the financial support.}}
\address{Institute for Communications Technology\\
	Technische Universit{\"a}t Braunschweig\\
	\{ziyue.zhao, s.elshamy, t.fingscheidt\}@tu-bs.de}
\begin{document}

\ninept
\maketitle

\begin{sloppy}

\begin{abstract}
\vspace{-0.1cm}
Single-channel speech enhancement with deep neural networks (DNNs) has shown promising performance and is thus intensively being studied. In this paper, instead of applying the mean squared error (MSE) as the loss function during DNN training for speech enhancement, we design a perceptual weighting filter loss motivated by the weighting filter as it is employed in analysis-by-synthesis speech coding, e.g., in code-excited linear prediction (CELP). The experimental results show that the proposed simple loss function improves the speech enhancement performance compared to a reference DNN with MSE loss in terms of perceptual quality and noise attenuation. The proposed loss function can be advantageously applied to an existing DNN-based speech enhancement system, without modification of the DNN topology for speech enhancement. 
\end{abstract}

\vspace{-0.5cm}
\section{Introduction}
\vspace{-0.1cm}
Single-channel speech enhancement aims to improve the quality and intelligibility of a speech signal degraded by additive noise, where the noisy mixture signal from only one microphone is available. 
As a widely researched problem, quite a number of contributions have been made over past decades, including conventional and data-driven speech enhancement approaches. 
Data-driven approaches have shown promising performance under non-stationary noise conditions, and they receive increasing research attention~\cite{Mirsamadi2016}. 
Therein, a regression for the spectral weighting rule indexed by the \textit{a posteriori} signal-to-noise ratio (SNR) and \textit{a priori} SNR for each frequency bin is trained and shows superior performance compared to conventional approaches~\cite{Fingscheidt2006datadriven, fingscheidt2008environment}.
As deep neural networks (DNNs) provide an effective way for supervised learning, DNN-based speech enhancement is intensively studied~\cite{wang2013towards,narayanan2013ideal,wang2014training,xu2015regression}. DNNs are trained to classify the noisy mixture signal into speech or noise for each frequency bin, known as ideal binary mask (IBM) \cite{wang2013towards}. Furthermore, some ratio masks are proposed as the targets for DNN training, e.g., the ideal ratio mask (IRM) \cite{narayanan2013ideal} and fast Fourier transform (FFT) mask \cite{wang2014training}. Those masks are usually trained directly by minimizing the error between the estimated mask and the oracle mask, which serves as a target. However, it is also possible to indirectly train the mask by introducing a multiplicative layer for the microphone signal during training, which allows to optimize the loss between clean speech targets and estimated clean speech. This has shown superior performance~\cite{erdogan2015phase}. % ,weninger2014discriminatively
Besides, DNNs are used to directly estimate the clean speech from the noisy speech by a regression~\cite{xu2015regression}. 

Loss functions play a key role in DNN training for speech enhancement, and the mean squared error (MSE) is a straightforward choice. 
The MSE loss is mostly applied in the same domain as the network's input~\cite{xu2015regression}, but it can be also computed in a different \mbox{domain} to exploit additional domain knowledge~\cite{pandey2018new}.
Instead of using the MSE loss, some other loss functions are designed to optimize the  perceptual metrics during the training, e.g., short-time objective intelligibility (STOI)~\cite{taal2011algorithm} and/or perceptual evaluation of speech quality (PESQ)~\cite{ITUT_pesq} are taken into account as loss functions. However, since the computation of both metrics is non-differentiable, some approximations have to be introduced in order to obtain fully differentiable loss functions for backpropagation~\cite{fu2017end,kolbcek2018monaural,zhao2018perceptually,martin2018deep,zhang2018training}. 
As an alternative, in~\cite{koizumi2018dnn}, the policy gradient method is applied on the basis of a sampling algorithm, leading to signals which achieve higher metric scores. Furthermore, in~\cite{fu2019metricgan}, the estimation of a metric score can be obtained by the discriminator of a generative adversarial network, and the generator then uses the prediction to decide for a gradient direction for optimization.
%In addition, the policy gradient method is applied on the basis of a sampling algorithm that the network is trained to increase the probability density function of the network output signals achieving high metric scores~\cite{koizumi2018dnn}. Besides, the estimation of a metric score can be obtained by the discriminator of the generative adversarial network, and the generator uses it to decide a gradient direction for optimization~\cite{fu2019metricgan}. 
Also, some perceptual weighting rules are proposed in loss functions for neural network training to achieve a better perceptual quality, e.g., high-energy frequency areas are emphasized in the loss functions in~\cite{xia2014wiener,liu2017perceptually}. However, a perceptual model is not considered in~\cite{xia2014wiener} and the loss function proposed in~\cite{liu2017perceptually} contains an empirical function to balance boosting high energy components and suppressing low energy components.
Besides, the absolute threshold of hearing and masking principles from psychoacoustics~\cite{painter1997review} are applied to construct the loss function in~\cite{kumar2016speech_interspeech}, where a slight improvement over using the MSE loss is found at very low SNR levels. 
In~\cite{han2016perceptual}, the authors propose a multi-target training that incorporates a perceptual weighting loss. However, the approach requires to change the network topology by adding additional output nodes, which results in a doubling of output nodes in~\cite{han2016perceptual}. This hinders the straightforward integration of the concept into existing frameworks as not only additional training targets have to be extracted, but also the topology has to be altered.

%\vspace{-0.1cm}
In this paper, two perceptual weighting filters from code-excited linear predictive (CELP) speech coding, as in, e.g., adaptive multi-rate (AMR)~\cite{AMR3GPP}, wideband AMR (AMR-WB)~\cite{AMRWB3GPP}, and enhanced voice services (EVS)~\cite{EVS3GPP}, are applied to design the loss function for DNN training in speech enhancement. The perceptual weighting filter is originally used to shape the coding noise\,/\,quantization error to be less audible by exploiting the masking property of the human ear~\cite{backstrom2017speech}. Motivated by the success of the weighting filter when used to shape coding noise, we apply it in this work to design the loss function to achieve improved perceptual quality in the context of speech enhancement, aiming to combat acoustic background noise (instead of coding noise, as has been done in~\cite{christoph2019learning}). We propose to extract the frequency response of the weighting filter from the clean speech and directly apply it to the loss function, resulting in an unaltered DNN architecture with no need to change the topology as in~\cite{han2016perceptual}. This results in an approach which is easy to integrate into existing frameworks. 

%\vspace{-0.1cm}
The paper is structured as follows: In Section 2 we briefly introduce the reference DNN-based speech enhancement with MSE loss. In Section 3 we describe the adopted weighting filter and how it is applied to the loss function in DNN training. Section 4 presents the experimental setup, the evaluation results, and
the discussion. Finally, some conclusions are drawn in Section 5. 

\vspace{-0.2cm}
\section{Reference DNN with MSE LOSS}
\label{sec_refDNN}
\vspace{-0.2cm}
\begin{figure}[tp]

	\psfrag{A}[cc][cc]{\footnotesize $S_{\ell}\!(k)$}
	\psfrag{B}[cc][cc]{\footnotesize $D_{\ell}\!(k)$}
	\psfrag{C}[cc][cc]{\footnotesize $Y_{\ell}\!(k)$}
	\psfrag{D}[cc][cc]{\footnotesize $|Y_{\ell}\!(k)|$}
	\psfrag{E}[cc][cl]{\footnotesize $[|Y_{\ell\!-\!2}\!(k)|,\!\ldots,\!|Y_{\ell\!+\!2}\!(k)|]$}
	\psfrag{F}[cc][cc]{\footnotesize $|M_\ell\!(k)|$}
	\psfrag{G}[cc][cc]{\footnotesize $\widehat{|S_\ell\!(k)|}$}
	\psfrag{H}[cc][cc]{\footnotesize $E_{\ell}\!(k)$}
	\psfrag{I}[cc][cc]{\footnotesize $E^{\text{w}}_{\ell}\!(k)$}
	\psfrag{J}[cc][cc]{\footnotesize $|S_\ell\!(k)|$}
	\psfrag{K}[cc][cc]{\footnotesize $a_\ell\!(k)$}
	\psfrag{L}[cc][cc]{\footnotesize $W_\ell\!(z)$}
	\psfrag{M}[cc][cc]{\footnotesize $W_\ell\!(k)$}
	\psfrag{N}[cc][cc]{\footnotesize $|W\!_\ell\!(k)|$}
	
	\psfrag{O}[cc][cc]{\footnotesize Delay\& }
	\psfrag{P}[cc][cc]{\footnotesize Reshape}
	\psfrag{Q}[cc][cc]{\footnotesize DNN}
	\psfrag{R}[cc][cc]{\footnotesize Scaling}
	\psfrag{S}[cc][cc]{\footnotesize $\left \| \cdot \right \|^2_2$} % $L_2$-Norm
	\psfrag{T}[cc][cc]{\footnotesize min}
	\psfrag{U}[cb][rb]{\footnotesize MSE}
	\psfrag{z}[cc][rb]{\footnotesize optimization}
	
	\psfrag{V}[cc][cc]{\footnotesize LP}
	\psfrag{W}[cc][cc]{\footnotesize Analysis}
	\psfrag{X}[cc][cc]{\footnotesize Weighting}
	\psfrag{Y}[cc][cc]{\footnotesize Filter}
	\psfrag{Z}[cc][cc]{\footnotesize $z\!\!\rightarrow \!\!e^{j\!\frac{2\pi k}{K}}$}
	
	\psfrag{a}[cc][cc]{\footnotesize Weighting filter amplitude response}
	%\psfrag{y}[cc][cc]{\footnotesize amplitude response}
	%\psfrag{x}[cc][cc]{\footnotesize computation}
	
	\psfrag{b}[cc][cc]{\footnotesize $\gamma_1$}
	\psfrag{c}[cc][cc]{\footnotesize $\gamma_2$}
	
	\psfrag{w}[cc][cc]{\footnotesize Enhanced speech spectral amplitudes}
	%\psfrag{v}[cc][cc]{\footnotesize spectral amplitudes}
	
	\psfrag{d}[cc][cc]{\footnotesize $s_{\ell}(n)$}
	\psfrag{e}[cc][cc]{\footnotesize FFT}
	
	\centering
	\includegraphics[width=0.48\textwidth]{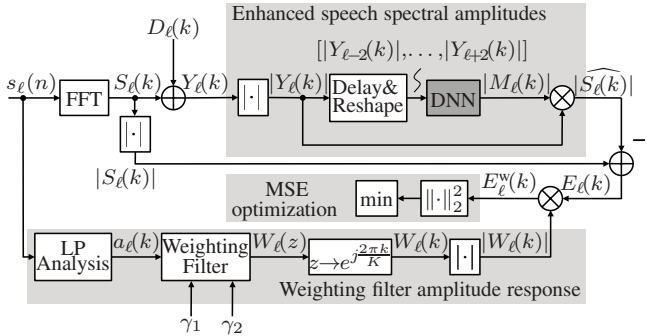}%{weighting_filter_loss_maskDNN_training_new.eps
	
	\vspace{-0.4cm}
	\caption{\textbf{Training stage} for DNN-based speech enhancement with perceptual weighting filter loss. For the reference with MSE loss just omit the lower branch and set $|W_{\ell}|\!=\!1$.}
	\label{fig_wfDNN_training}
	\vspace{-0.3cm}
\end{figure}

In this paper, the single-channel noisy mixture signal is modelled as $y(n)\!=\!s(n)\!+\!d(n)$, with $s(n)$ being the clean speech signal, $d(n)$ being the additive noise, and $n$ being the discrete-time sample index. By applying a $K$-point FFT, the frequency domain analogue is $Y_{\ell}(k)\!=\!S_{\ell}(k)\!+\!D_{\ell}(k)$, with $\ell$ being the frame index and $k\!\in\left \{ 0,\!1,\!\ldots\!,\!K\!-\!1 \right \}$ being the frequency bin index. The frames are assembled by applying the periodic Hann window function with \mbox{16 ms} frame length and 50\% overlap. 

The reference DNN trained with MSE loss and the proposed approach using the weighting filter loss share the same enhancement structure apart from the loss function. The shared structure is shown in the upper grey-shaded part of \figref{fig_wfDNN_training} (denoted as ``Enhanced speech spectral amplitudes''). The input of the DNN are the spectral amplitudes of the current noisy signal's frame $|Y_{\ell}(k)|$ along with the spectral amplitudes of two left and two right context frames, resulting in the number of DNN input nodes being $5\!\times\!129\!=\!645$. Note that only the first $K\!/2\!+\!1=\!129$ spectral amplitude values per frame are to be enhanced with $K\!=\!256$, since the other values can be obtained symmetrically. Then, the mask $|M_{\ell}(k)|$, implicitly predicted by the DNN, is multiplied to the noisy spectral amplitudes $|Y_{\ell}(k)|$ to generate the enhanced speech spectral amplitudes 
\vspace{-0.2cm}
\begin{equation}
\label{equ_enhan_mask}
\widehat{|S_\ell(k)|}=|Y_{\ell}(k)|\cdot |M_{\ell}(k)|.
\vspace{-0.1cm}
\end{equation} 
It is worth noting that the targets for DNN training are the clean speech spectral amplitudes $|S_\ell(k)|$. Finally, the enhanced speech signal $\hat{s}(n)$ is obtained by performing the inverse FFT together with the noisy phase from $Y_{\ell}(k)$ and overlap-add synthesis. 

The DNN topology used in this paper is shown in \figref{fig_dnn_topo}. The DNN contains five hidden layers, which are fully-connected layers, followed by batch normalization, leaky rectified linear unit (ReLU) activation functions, and dropout with a dropout rate of 0.2. In addition, skip connections are used to add up the layer outputs with matching dimensions, in order to ease the vanishing gradient problem during training~\cite{he2016deep}. The output layer, implicitly predicting the mask spectral amplitudes $|M_{\ell}(k)|$, has $K\!/2\!+\!1=\!129$ nodes with sigmoid activation functions to limit the output to $0\!\leq\! |M_{\ell}(k)|\!\leq\!1$. 
%The second layer output is added to the third layer output as the forth layer input and this sum is added to the forth layer output as the fifth layer input. Batch normalization is used for all layers except for the input layer. By doing so, the vanishing gradient problem during training can be eased. 

\begin{figure}[tp]
	
	\psfrag{a}[cc][cc]{\footnotesize $[\,|Y_{\ell\!-\!2}\!(k)|,$}
	\psfrag{z}[cc][cc]{\footnotesize $\ldots,$}
	\psfrag{y}[cc][cc]{\footnotesize $|Y_{\ell\!+\!2}\!(k)|\,]$}
	\psfrag{b}[cc][cc]{\footnotesize \rotatebox{90}{FC(1024)}}
	\psfrag{c}[cc][cc]{\footnotesize\rotatebox{90}{FC(512)}}
	\psfrag{d}[cc][cc]{\footnotesize \rotatebox{90}{FC(512)}}
	\psfrag{e}[cc][cc]{\footnotesize \rotatebox{90}{FC(512)}}
	\psfrag{f}[cc][cc]{\footnotesize \rotatebox{90}{FC(256)}}
	
	\psfrag{g}[cc][cc]{\footnotesize Output}
	\psfrag{i}[cc][cc]{\footnotesize Layer}
	\psfrag{j}[cc][cc]{\footnotesize $|M_{\ell}(k)|$}
	
	\psfrag{h}[cc][cc]{\footnotesize Hidden layers}
	\psfrag{k}[cr][cr]{\footnotesize DNN}
	
	\centering
	\includegraphics[width=0.475\textwidth]{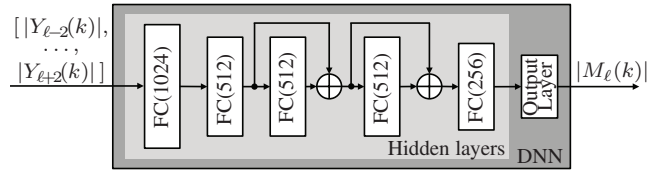}%
	
	\vspace{-0.3cm}
	\caption{Detailed view of the \textbf{DNN}: The operation FC($N$) stands for a fully-connected layer with $N$ nodes, followed by batch normalization, leaky ReLU activation functions, and dropout. The output layer contains batch normalization, a fully-connected layer with 129 nodes, and sigmoid activation functions.} % The output layer contains batch normalization, fully-connected layer with 192 nodes, and sigmoid activation function. 
	\label{fig_dnn_topo}
	\vspace{-0.2cm}
\end{figure}

\vspace{-0.2cm}
\section{Perceptual Weighting Filter Loss}
\label{sec_percep_weight}
\vspace{-0.2cm}
\subsection{Perceptual Weighting Filter in CELP Speech Coding}
\label{subsec_percep_weight_celp_coding}
In AMR speech coding, the perceptual weighting filter, employed in the analysis-by-synthesis search of the codebooks, is expressed according to~\cite{AMR3GPP} as
\vspace{-0.3cm}
\begin{equation}
\label{equ_wgh_filt}
W_{\ell}(z)=\frac{1-A_{\ell}(z/\gamma_1)}{1-A_{\ell}(z/\gamma_2)},
\vspace{-0.2cm}
\end{equation}
where $A_{\ell}(z/\gamma)\!=\!\sum_{i=1}^{N_p}\!a_{\ell}(i)\gamma^iz^{-i}$, $a_{\ell}(i)$ are the linear prediction (LP) coefficients of frame $\ell$, $N_p$ is the prediction order, and $\gamma_1$, $\gamma_2$ are the perceptual weighting factors. As the search of the codebooks is done by minimizing the weighted error between the clean speech and the coded speech, the weighted error becomes spectrally white, meaning that the final (unweighted) coding error is proportional to the \textit{inverse} weighting filter $1/W_{\ell}(z)$.
%follow the spectral shape of the \textit{inverse} weighting filter $1/W_{\ell}(z)$. 
This inverse weighting filter has similarities to the structure of the clean speech spectral envelope which exploits the masking property of the human ear: More energy of the quantization error will be in the speech formant regions, as $1/W_{\ell}(z)$ is somewhat below the spectral envelope there. 

A variant of the above weighting filter, originally used in AMR-WB (also used in the recent EVS codec)~\cite{AMRWB3GPP,EVS3GPP} is proposed as
\vspace{-0.1cm}
\begin{equation}
\label{equ_wgh_filt_wb}
W'_{\ell}(z)=1-A'_{\ell}(z/\gamma_1),
\vspace{-0.1cm}
\end{equation}
where the LP coefficients $a'_{\ell}(i)$ are computed based on the speech being preemphasized by a filter $H_{\text{pre}}(z)\!=\!1\!-\!\beta z^{-1}$. 
Note that the weighting filter in the codecs is originally of the form $\big(1\!-\!A'_{\ell}(z/\gamma_1)\big)/H_{\text{pre}}(z)$ and a de-emphasis filter $H_{\text{pre}}^{-1}(z)$ is applied in the decoder. Therefore, the quantization error is actually proportional to $1/\big(1\!-\!A'_{\ell}(z/\gamma_1)\big)$~\cite{AMRWB3GPP}, resulting in the equivalent weighting filter (\ref{equ_wgh_filt_wb}).
Both filter types being employed as potential weighting filter losses will be evaluated and discussed in the next section. % on the development dataset 

To intuitively understand this shaping effect, we exemplarily show the clean speech spectral envelope, the pedestrian (PED) noise amplitude spectrum, and the noise shaped by the inverse weighting filter (2) for an example frame in \figref{fig_invers_wf}. This example frame is from the speech file \texttt{bbaf1s} of the Grid corpus~\cite{cooke2006audio} downsampled to 16 kHz, the pedestrian noise is from the CHiME-3 dataset~\cite{barker2015third}, and the SNR is 5 dB. The AMR weighting filter (2) is used in this example with $N_p\!=\!16$, $\gamma_1\!=\!0.92$, and $\gamma_2\!=\!0.6$.
It shows nicely, how the noise is shaped to be ``hidden" under the speech spectral envelope, so that the shaped noise will be less perceivable by the human ear. 
% (\textcolor{myblue}{blue dashed line}) (\textcolor{myblue}{blue solid line}) 

\begin{figure}[tp]
	
	\psfrag{0}[tl][tl]{\footnotesize $0$}
	\psfrag{16}[tc][tc]{\footnotesize $16$}
	\psfrag{32}[tc][tc]{\footnotesize $32$}
	\psfrag{48}[tc][tc]{\footnotesize $48$}
	\psfrag{64}[tc][tc]{\footnotesize $64$}
	\psfrag{80}[tc][tc]{\footnotesize $80$}
	\psfrag{96}[tc][tc]{\footnotesize $96$}
	\psfrag{112}[tc][tc]{\footnotesize $112$}
	\psfrag{128}[tc][tc]{\footnotesize $128$}
	
	\psfrag{45}[cr][cr]{\footnotesize $45$}
	\psfrag{20}[cr][cr]{\footnotesize $20$}
	\psfrag{30}[cr][cr]{\footnotesize $30$}
	\psfrag{40}[cr][cr]{\footnotesize $40$}
	\psfrag{50}[cr][cr]{\footnotesize $50$}
	\psfrag{55}[cr][tr]{\footnotesize $55$}
	\psfrag{25}[cr][cr]{\footnotesize $25$}
	\psfrag{35}[cr][cr]{\footnotesize $35$}
	
	\psfrag{ylabeltext}[bc][tc]{\footnotesize $10$log$_{10}(\cdot)$ [dB]}
	\psfrag{Frequencybinsk}[tc][bc]{\footnotesize Frequency bin $k$}
	
	%	\psfrag{Clean-speech-spectral-envelopeeee}[tl][tl]{\scriptsize Clean speech spectral envelope: $|S_\ell(k)|$}
	%	\psfrag{PED-noise-spectrum}[tl][tl]{\scriptsize PED noise spectrum: $|N_\ell(k)|$}
	%	\psfrag{Inverse-weighted-PED-noise-spectrummmm A}[cl][cl]{\scriptsize Inverse weighted PED noise spectrum: $\frac{|N_\ell(k)|}{|W_\ell(k)|}$}

	\centering
	\includegraphics[width=0.45\textwidth]{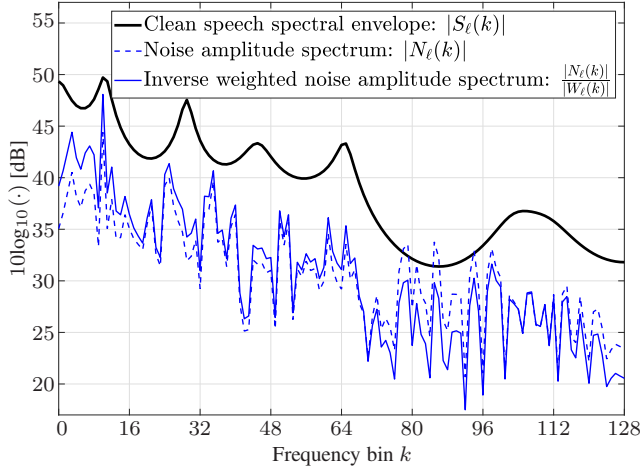}   %   ,angle=270  width=0.5\textwidth  height=\columnwidth
	\vspace{-0.1cm}
	\caption{Clean speech spectral envelope, pedestrian (PED) noise amplitude spectrum, and the inverse weighted PED noise amplitude spectrum for an example frame. }
	\label{fig_invers_wf}
	\vspace{-0.2cm}
\end{figure}

\vspace{-0.4cm}
\subsection{Perceptual Weighting Filter Loss in DNN Training}
\label{subsec_percep_weight_dnn_train}
Now that the perceptual weighting filter from CELP speech coding has been revisited, we show how a weighting filter loss is constructed for DNN training in \figref{fig_wfDNN_training}. We derive the loss from the AMR weighting filter, while the AMR-WB weighting filter is obtained straightforward. First, LP analysis is performed based on the clean speech frame $s_{\ell}(n)$ to obtain the frame-wise LP coefficients $a_{\ell}(i)$. Then, the weighting filter $W_{\ell}(z)$ is calculated by (\ref{equ_wgh_filt}) and the corresponding frequency amplitude response is obtained as %  with the order $N_p\!=\!16$
\vspace{-0.1cm}
\begin{equation}
\label{equ_wgh_filt_z_k}
W_{\ell}(k)=W_{\ell}(z)\!\Bigm|_{z=e^{j\!\frac{2\pi k}{K}}},\ \ \ \ \  k\in\{0,1,\ldots,K/2\}.
\vspace{-0.1cm}
\end{equation}
Finally, the loss function is obtained as
\vspace{-0.25cm}
\begin{equation}
\label{equ_wgh_filt_loss}
%\begin{split}
%%J_{\ell}&=\frac{1}{|\mathcal{K}|} \sum_{k\in\mathcal{K}}\big(E_{\ell}^{\text{w}}(k)\big)^2\\
%%&=\frac{1}{|\mathcal{K}|} \sum_{k\in\mathcal{K}}\big(|W_{\ell}(k)|\cdot (|S_{\ell}(k)|-\widehat{|S_{\ell}(k)|})\big)^2,
%J_{\ell}= \sum_{k\in\mathcal{K}}\big(E_{\ell}^{\text{w}}(k)\big)^2= \sum_{k\in\mathcal{K}}\big( |W_{\ell}(k)|\cdot E_{\ell}(k)\big)^2
J_{\ell} =  \big(E^{\text{w}}_{\ell}(0)\big)^2+\big(E^{\text{w}}_{\ell}(\tfrac{K}{2})\big)^2+2\cdot\!\!\sum_{k=1}^{K\!/2\!-\!1}\big(E_{\ell}^{\text{w}}(k)\big)^2,
%=  \big(|W_{\ell}(0)|\cdot E_{\ell}(0)\big)^2+\big(|W_{\ell}(\tfrac{K}{2})|\cdot E_{\ell}(\tfrac{K}{2})\big)^2+2\cdot\sum_{k\in\mathcal{K}}\big( |W_{\ell}(k)|\cdot E_{\ell}(k)\big)^2 
%\end{split}
\vspace{-0.2cm}
\end{equation}
where $E_{\ell}^{\text{w}}(k)\!=\!|W_{\ell}(k)|\cdot E_{\ell}(k)$ is the weighted error, and  $E_{\ell}(k)\!=\!|S_{\ell}(k)|-\widehat{|S_{\ell}(k)|}$ is the training error.
%As the loss function is minimized during the training stage, the training error $E_{\ell}(k)$ is weighted by the inverse weighting filter $1/|W_{\ell}(k)|$: A similar consequence as in CELP speech coding described in Section \ref{subsec_percep_weight_celp_coding}. The quantization error during the search in the codebook is weighted by the inverse weighting filter. 
As $E_{\ell}^{\text{w}}(k)$ becomes spectrally white, we find that $E_{\ell}(k)\!\sim\!1/|W_{\ell}(k)|$.
As a result, the residual noise is expected to be less audible. It is worth noting that in the enhancement stage, the inference process of this DNN is the same as for the reference DNN.

\vspace{-0.2cm}
\section{Experimental Evaluation}
\vspace{-0.1cm}
\label{sec_exp_eval}
\subsection{Experimental Setup and Evaluation Metrics}
\label{subsec_exp_setup}
\vspace{-0.1cm}
The speech material used in this paper is from the Grid corpus~\cite{cooke2006audio} and is downsampled to 16 kHz. We randomly select 8 female and 8 male speakers, and each speaker includes 100 sentences with a duration of 3 seconds per sentence. Thus, the total amount of clean speech used for training and validation is 80 minutes long. Another 2 female and 2 male speakers with 20 sentences each are used for testing. 
The additive background noise is from the CHiME-3 dataset: Pedestrian noise (PED), caf\'{e} noise (CAF), and street noise (STR) are used for the training, validation, and test, while bus noise (BUS) is used additionally as an unseen noise type for testing. Training  and validation material contains the noisy data at six SNR levels (from -5 dB to 20 dB with 5 dB step size), and these two sets are obtained by a split with a ratio of 4\,$:$\,1 with a proportional number of sentences covered by all the speakers, noise types and SNR levels. Note that the noise material used in the test set is also disjoint from that for training and validation.

Regarding the DNN training, the input is first normalized to have zero mean and unit variance based on the statistics of the training data. Then, the weights of the DNN are trained by minimizing the loss function, applying the Adam algorithm with the learning rate being $5\!\cdot\!10^{-4}$. The weights are updated after each minibatch containing 128 input and target pairs, which are randomly selected from the training data. 
%~\cite{Kingma2015adam} 

% EVALUATION for different filter coefficient
\begin{figure}[tp]
	
	\psfrag{01}[cc][tc]{\scriptsize $1$ } 
	\psfrag{0098}[cc][tc]{\scriptsize $0.98$}
	\psfrag{0096}[cc][tc]{\scriptsize $0.96$}
	\psfrag{0094}[cc][tc]{\scriptsize $0.94$}
	\psfrag{0092}[cc][tc]{\scriptsize $0.92$}
	\psfrag{009}[cc][tc]{\scriptsize $0.9$}
	\psfrag{0088}[cc][tc]{\scriptsize $0.88$}
	\psfrag{0086}[cc][tc]{\scriptsize $0.86$}
	\psfrag{0084}[cc][tc]{\scriptsize $0.84$}
	\psfrag{0082}[cc][tc]{\scriptsize $0.82$}
	\psfrag{008}[cc][tc]{\scriptsize $0.8$}
	\psfrag{0078}[cc][tc]{\scriptsize $0.78$}
	\psfrag{0076}[cc][tc]{\scriptsize $0.76$}
	\psfrag{0074}[cc][tc]{\scriptsize $0.74$}
	\psfrag{0072}[cc][tc]{\scriptsize $0.72$}
	\psfrag{007}[cc][tc]{\scriptsize $0.7$}
	
	\psfrag{2.1}[cl][tl]{\footnotesize $2.1$}
	\psfrag{2.15}[cr][cr]{\footnotesize $2.15$}
	\psfrag{2.2}[cr][cr]{\footnotesize $2.2$}
	\psfrag{2.25}[cr][cr]{\footnotesize $2.25$}
	\psfrag{2.3}[cr][cr]{\footnotesize $2.3$}
	\psfrag{2.35}[cr][cr]{\footnotesize $2.35$}
	\psfrag{2.4}[cr][cr]{\footnotesize $2.4$}
	\psfrag{2.45}[cr][cr]{\footnotesize $2.45$}
	\psfrag{2.5}[cr][cr]{\footnotesize $2.5$}
	\psfrag{2.55}[cr][cr]{\footnotesize $2.55$}
	
	\psfrag{ylabeltext}[bc][tc]{\footnotesize PESQ}
	\psfrag{BaselineDNNNN}[cc][cc]{\footnotesize Reference DNN}
	
	\psfrag{AMR}[cc][cc]{\footnotesize AMR}
	\psfrag{AMRWB}[cc][cc]{\footnotesize AMR-WB}
	\psfrag{ggaamma1}[cc][cc]{\footnotesize $\gamma_1$}
	\psfrag{ggaamma12}[cc][cc]{\footnotesize $\gamma_1$}
	
	\centering
	\centerline{\includegraphics[width=0.45\textwidth]{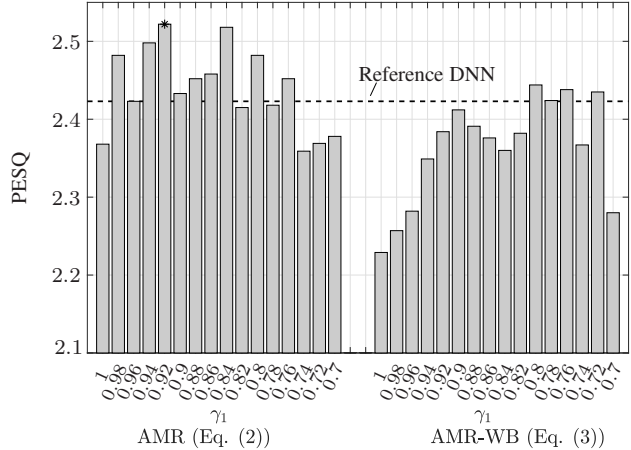}}   %  fig_elav_data  ,angle=270  width=0.5\textwidth  height=\columnwidth
	\vspace{-0.3cm}
	\caption{PESQ performance for the proposed loss functions applying the weighting filters from AMR and AMR-WB, with the perceptual weighting factor $\gamma_1\!\in\!\{1,\!0.98,\!\ldots,\!0.7\}$, $\gamma_2\!=\!0.6$ (AMR), and the preemphasis filter factor $\beta\!=\!0.68$ (AMR-WB) on the \textit{development} dataset. The optimal setting is marked with $\bm{\ast}$.}
	\label{fig_eval_data}
	\vspace{-0.3cm}
\end{figure}

In order to evaluate the speech enhancement system including both, speech and noise components, PESQ~\cite{ITUT_pesq_wb_corri} and STOI~\cite{taal2011algorithm} are used to measure the quality and the intelligibility of the enhanced speech. Also, SNR improvement (SNRI)~\cite{ITUT_g160_snri} is used to measure the SNR improvement achieved by the speech enhancement system. 
%Note that the PESQ is measured based on the two-sentence files, generated by simply concatenating two single-sentence files, 

Regarding the evaluation metric for the speech component, segmental speech-to-speech-distortion ratio (SSDR) is calculated after~\cite{elshamy2017instantaneous} as
\vspace{-0.1cm}
\begin{equation}
\label{equ_ssdr}
\text{SSDR}=\frac{1}{\left|\mathcal{L}\right|}\sum_{\ell\in\mathcal{L}}\text{SSDR}(\ell),
\vspace{-0.1cm}
\end{equation} 
where $\mathcal{L}$ is the set of frame indices for speech active frames and $\text{SSDR}(\ell)$ is limited from $R_{\text{min}}\!=\!-10$ dB to $R_{\text{max}}\!=\!30$ dB by $\text{SSDR}(\ell)\!=\!\text{max}\left \{ \text{min}\left \{ \text{SSDR}'(\ell),\!R_{\text{max}} \right \}\!,\! R_{\text{min}} \right \}$. The term $\text{SSDR}'(\ell)$ is actually calculated as 
\vspace{-0.2cm}
\begin{equation}
\label{equ_ssdr_prim}
\text{SSDR}'(\ell)=10\text{log}_{10}\left [ \frac{\sum_{n\in\mathcal{N}_{\ell}}s(n)^2}{\sum_{n\in\mathcal{N}_{\ell}}\left ( \tilde{s}(n)-s(n) \right )^2} \right ] \; [\text{dB}],
\vspace{-0.1cm}
\end{equation}
where $\mathcal{N}_{\ell}$ is the set of sample indices $n$ belonging to frame $\ell$, $s(n)$ is the clean speech, and $\tilde{s}(n)$ is the time-aligned \textit{filtered} speech component. Note that this \textit{filtered} speech component $\tilde{s}(n)$ and the following \textit{filtered} noise component $\tilde{d}(n)$ are obtained on the basis of (\ref{equ_enhan_mask}), but replacing the noisy speech spectral amplitudes $|Y_{\ell}(k)|$ by either the speech component spectral amplitudes $|S_{\ell}(k)|$, or the noise component spectral amplitudes $|D_{\ell}(k)|$, respectively~\cite{gustafsson1996optimization}. 
%output of the speech enhancement system with the input being \textit{clean} speech component. 

Another evaluation metric we use in this paper to measure the SNR improvement is $\Delta\text{SNR}\!=\!\text{SNR}_{\text{out}}\!-\!\text{SNR}_{\text{in}}$, where $\text{SNR}_{\text{in}}$ is the SNR level for the noisy mixture $y(n)$, and $\text{SNR}_{\text{out}}$ is calculated based on $\tilde{s}(n)$ and $\tilde{d}(n)$.
%, with $\tilde{d}(n)$ being the output of the speech enhancement system using \textit{noise} component as input. 

%\vspace{-0.7cm}
\subsection{Experimental Results and Discussion}
\label{subsec_exp_res_dis}
\vspace{-0.1cm}

% TEST table
\begin{table}[tp]
	\begin{center}
		\scriptsize
		\setlength\tabcolsep{1.9pt} % default value: 6pt
		\begin{tabular}{ m{0.54cm}<{\centering} | m{1.7cm}<{\centering} | m{1.22cm}<{\centering} | m{1.22cm}<{\centering} |  m{0.75cm}<{\centering} m{0.65cm}<{\centering}  m{1.2cm}<{\centering} m{0.0001cm}<{\centering} }
			\cline{1-7}
			\multirow{2}{*}{\stz{Noise}} & \multirow{3}{*}{\stz{Method}} & \stz{Noise}&  \stz{Speech}&  \multicolumn{3}{c}{\multirow{2}{*}{Total}} & \\ [1.6pt]
			\multirow{2}{*}{Type}& & Component & Component  &  &\multicolumn{1}{c}{} & \multicolumn{1}{c}{} & \multicolumn{1}{c}{} \\
			\cline{3-7}
			& & \stz{$\Delta$SNR [dB]} & \stz{SSDR [dB]}  & \stz{PESQ} & \stz{STOI} & \stz{SNRI [dB]} & \\[1pt]
			\cline{1-7}
			
			 \multirow{3}{*}{PED} & \stz{Noisy} & - & - & \stz{1.98} & \stz{0.66} & - & \\
			 & Reference DNN & 4.78 & \bf 17.34  & 2.50 &\bf 0.69 & 10.56 & \\
			& Weighted DNN & \bf 5.55 &  15.57 & \bf 2.60 & \bf 0.69 & \bf 14.48 & \\
			\cline{1-7}
			
			\multirow{3}{*}{CAF}	& \stz{Noisy} & - & - & \stz{1.99} & \stz{0.64} & - & \\
			 & Reference DNN & 4.84 & \bf 17.43  & 2.50 & \bf 0.68 & 11.04 & \\
			& Weighted DNN & \bf 5.48 & 15.79 & \bf 2.57 & \bf 0.68 & \bf 15.82 & \\
			\cline{1-7}
			
			\multirow{3}{*}{STR} & \stz{Noisy} & - & - & \stz{1.97} & \stz{0.69} & - & \\
			 & Reference DNN & 5.77 & \bf 18.54  &  2.52 & \bf 0.72 & 12.11 & \\
			& Weighted DNN & \bf 6.41 & 16.99 & \bf 2.63 & \bf 0.72 & \bf 16.47 & \\
			\cline{1-7}
			
			\stz{BUS} & \stz{Noisy} & - & - & \stz{2.19} & \stz{0.72} & - & \\
			 (un-& Reference DNN & 4.03 & \bf 20.85  &  2.65 & \bf 0.74 & 6.08 & \\
			seen)& Weighted DNN & \bf 4.93 & 19.76 & \bf 2.70 & \bf 0.74 & \bf 9.71 & \\
			\cline{1-7}

		\end{tabular}
	\end{center}
	
	\vspace{-0.4cm}
	\caption{Evaluation metrics for the proposed approach (denoted as weighted DNN) and the reference DNN approach under different noise conditions. $\Delta$SNR, SSDR, and SNRI are measured in dB. The better approach is in \textbf{boldface}.}
	\label{tab_test}
	\vspace{-0.4cm}
\end{table}

We first search for the optimal parameter $\gamma_1$ of the perceptual weighting filter applied to the loss function by conducting experiments on a development dataset. This development dataset is a subset of the validation data, which uses a quarter of the data from the validation dataset covering all speakers, noise types, and SNR levels. It is used to decrease the amount of data for optimal parameter search, thus improving efficiency. As shown in \figref{fig_eval_data}, the two abovementioned perceptual weighting filters (see (\ref{equ_wgh_filt}) and (\ref{equ_wgh_filt_wb}), with $N_p\!=\!16$) are evaluated with various perceptual weighting factors. 
Regarding the weighting filter from AMR, we search the optimal $\gamma_1$, as different $\gamma_1$ values are also applied for different bitrates in the AMR codec~\cite{AMR3GPP}. In the meanwhile, we keep $\gamma_2\!=\!0.6$ unchanged because an informal search on $\gamma_2$ with limited values shows that  the performance has a rather weak dependency on $\gamma_2$.
The same search range of $\gamma_1$ is also investigated for the weighting filter from AMR-WB, and we keep the preemphasis filter factor $\beta\!=\!0.68$ as in~\cite{AMRWB3GPP}. The factor combinations actually define the spectral shape and the spectral tilt of the weighting filter. 
It can be seen that under the investigated parameters, the weighting filters with various $\gamma_1$ from AMR shows generally better performance compared to those from AMR-WB.
%, when they are applied to the loss function for DNN training. 
The weighting filters of AMR with $\gamma_1$ in the selected range (left part in \figref{fig_eval_data}) can outperform the reference DNN except for some outlier values of $\gamma_1$.
The optimal settings are the weighting filter from AMR (\ref{equ_wgh_filt}) with $\gamma_1\!=\!0.92,\,\gamma_2\!=\!0.6$. The DNN model optimized by the loss function with the optimal weighting filter settings will be used in the following experiments (denoted as Weighted DNN).

In \tabref{tab_test}, we report the results on the test dataset and compare them to the reference DNN. The results are collected in four noise conditions and all values are averaged over six SNRs from \mbox{-5} to 20 dB. 
We can see the PESQ score improvements from the noisy speech $y(n)$ to the enhanced speech $\hat{s}(n)$ for both approaches. Compared to the reference DNN, the proposed method achieves improved PESQ performance in the range of 0.07...0.11 points for the three seen noise types (PED, CAF, and STR), and 0.05 points improvement for the unseen BUS noise condition. The STOI measures are comparable for the two approaches. 
Regarding the SNR improvement, the proposed approach consistently outperforms the reference DNN in terms of SNRI and $\Delta$SNR by more than 3.5 dB and 0.6 dB, respectively, for all four noise conditions. 
The results support the effectiveness of the proposed loss function, showing that the difference between the enhanced speech and the clean speech is less perceptually significant.
%the SNRI and $\Delta$SNR both indicate that the proposed approach can substantially attenuate the noise more effectively under all four noise conditions.

We notice that the SSDR is decreased by using the weighting filter loss function, which is easy to explain: The trained DNN with the weighting filter loss is biased to focus on the spectrum, i.e., more focus is put on the clean speech spectral valley regions and less on the formant regions. This effectively improves the perceptual quality of the enhanced speech as shown in the above PESQ scores, however, it introduces some measurable distortions to the speech component. This proves that the weighting filter loss does what it is expected to do: It does not excel the reference DNN in terms of MSE (or SSDR), but only subjectively (shown by PESQ).
%A further idea to alleviate the distortion could be to add another weighting filter loss function based on the clean speech spectral amplitudes and the speech component spectral amplitudes. 
%The weighting filter is applied to the loss, which is based on the clean speech spectral amplitudes $|S_{\ell}(k)|$ and the enhanced speech spectral amplitudes $|\widehat{S_{\ell}(k)|}$ and thus     The trained DNN with the weighting filter loss have the biased focus on the spectrum, i.e., more focus has been put on the clean speech spectral valley regions and less on the formant regions. This introduces some distortions to the speech component, however, This is perceptually efficient from the above discussions regarding the PESQ scores. However, distortions are therefore introduced when the input of the enhancement system are clean speech spectral amplitudes $|S_{\ell}(k)|$ instead of the noisy speech spectral amplitude $|Y_{\ell}(k)|$.
%as in the training stage
%by the weighting filter loss (compared to the normal MSE loss) to shape the residual noise (difference between enhanced speech $\hat{s}(n)$ and the clean speech $s(n)$) to be less audible. 

We further show some results in various SNR levels in \figref{fig_test_data_curve}, where all the values are averaged over the four noise types. The PESQ score improvement for the proposed approach over the reference DNN is more significant in higher SNR levels than in lower SNR levels. As already shown in \tabref{tab_test}, the STOI measures for the two approaches are quite comparable also across various SNR levels. Regarding the SNRI metric, the proposed method clearly excels the reference DNN in all SNR conditions, obtaining 4.17 dB SNRI improvement on average of SNR conditions and noise types.

% TEST for different metrics
%\vspace{-0.2cm}
\begin{figure}[t!]
	\psfrag{AMR1}[cc][cl]{\scriptsize AMR, $\gamma_1\!=\!1$ } % 
	\psfrag{AMR255}[cc][cl]{\scriptsize AMR, $\gamma_1\!=\!0.98$}
	\psfrag{AMR355}[cc][cl]{\scriptsize AMR, $\gamma_1\!=\!0.96$}
	\psfrag{AMR455}[cc][cl]{\scriptsize AMR, $\gamma_1\!=\!0.94$}
	\psfrag{AMR555}[cc][cl]{\scriptsize AMR, $\gamma_1\!=\!0.92$}
	\psfrag{AMR65}[cc][cl]{\scriptsize AMR, $\gamma_1\!=\!0.9$}
	\psfrag{EVS1}[cc][cl]{\scriptsize EVS, $\gamma_1\!=\!1$}
	\psfrag{EVS255}[cc][cl]{\scriptsize EVS, $\gamma_1\!=\!0.98$}
	\psfrag{EVS355}[cc][cl]{\scriptsize EVS, $\gamma_1\!=\!0.96$}
	\psfrag{EVS455}[cc][cl]{\scriptsize EVS, $\gamma_1\!=\!0.94$}
	\psfrag{EVS555}[cc][cl]{\scriptsize EVS, $\gamma_1\!=\!0.92$}
	\psfrag{EVS65}[cc][cl]{\scriptsize EVS, $\gamma_1\!=\!0.9$}
	
	\psfrag{1}[tl][tr]{\footnotesize $-5$}
	\psfrag{2}[tc][tc]{\footnotesize $0$}
	\psfrag{3}[tc][tc]{\footnotesize $5$}
	\psfrag{4}[tc][tc]{\footnotesize $10$}
	\psfrag{5}[tc][tc]{\footnotesize $15$}
	\psfrag{6}[tc][tc]{\footnotesize $20$}
	
	\psfrag{a}[cr][cl]{\footnotesize $5$}
	\psfrag{10}[cr][cr]{\footnotesize $10$}
	\psfrag{15}[cr][cr]{\footnotesize $15$}
	\psfrag{20}[cr][cr]{\footnotesize $20$}
	
	\psfrag{0.5}[cr][cr]{\footnotesize $0.5$}
	\psfrag{0.6}[cr][cr]{\footnotesize $0.6$}
	\psfrag{0.7}[cr][cr]{\footnotesize $0.7$}
	\psfrag{0.8}[cr][cr]{\footnotesize $0.8$}
	\psfrag{0.9}[cr][cr]{\footnotesize $0.9$}
	
	\psfrag{1.5}[cr][cr]{\footnotesize $1.5$}
	\psfrag{b}[cr][cr]{\footnotesize $2$}
	\psfrag{2.5}[cr][cr]{\footnotesize $2.5$}
	\psfrag{c}[cr][cr]{\footnotesize $3$}
	\psfrag{3.5}[cr][cr]{\footnotesize $3.5$}
	\psfrag{d}[cr][cr]{\footnotesize $4$}

	\psfrag{PESQMOS}[bc][tc]{\footnotesize PESQ}
	\psfrag{STOI}[bc][tc]{\footnotesize STOI}
	\psfrag{SNRIdB}[bc][tc]{\footnotesize SNRI [dB]}
	\psfrag{SNRdB}[tc][bc]{\footnotesize SNR [dB]}
	
	\psfrag{WeightedDNNNNN}[cc][cc]{\footnotesize Weighted DNN}
	\psfrag{BaselineDNN}[cl][cl]{\footnotesize Reference DNN}
	
	\psfrag{PESQtext}[tl][tl]{\footnotesize PESQ (noisy)}
	\psfrag{STOItext}[tl][tl]{\footnotesize STOI (noisy)}
	
	\centering
	\vspace{-0.1cm}
	\includegraphics[width=0.45\textwidth]{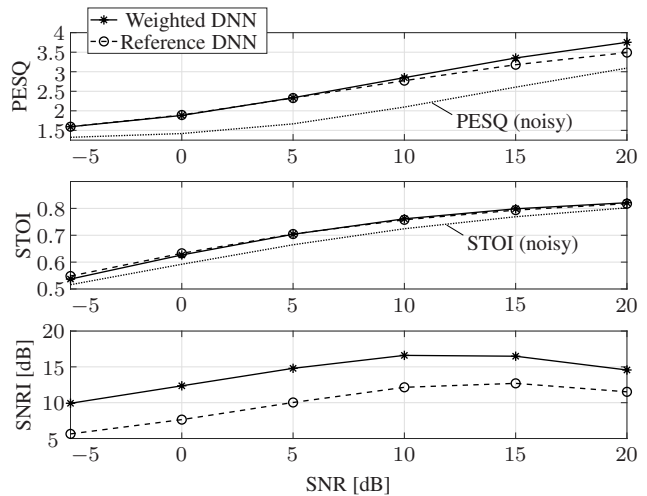}   %   ,angle=270  width=0.5\textwidth  height=\columnwidth
	\vspace{-0.1cm}
	\caption{PESQ, STOI, and SNRI measures for various SNRs averaged over all four noise types of the proposed approach (denoted as weighted DNN) and the reference DNN approach.}
	\vspace{-0.4cm}
	\label{fig_test_data_curve}
\end{figure}

\vspace{-0.4cm}
\section{Conclusions}
\label{sec_conc}
\vspace{-0.2cm}
In this paper, a perceptual weighting filter loss is designed for deep neural network (DNN) training in speech enhancement by applying the weighting filter from CELP speech coding. Simulation results show that the proposed approach outperforms the reference DNN trained with mean squared error loss in terms of speech quality measured by PESQ and significantly higher noise attenuation measured by more than 4 dB SNR improvement and 0.7 dB $\Delta$SNR on average over SNR levels and noise types.
The proposed loss function could be applied advantageously to an existing DNN-based speech enhancement system, without modification of the DNN topology and the speech enhancement framework. The source code for the proposed approach is available at \href{https://github.com/ifnspaml/Perceptual-Weighting-Filter-Loss}{https://github.com/ifnspaml/Perceptual-Weighting-Filter-Loss}. %\href{https://github.com/ifnspaml/Perceptual-Weighting-Filter-Loss}{https://github.com/ifnspaml/Perceptual-Weighting-Filter-Loss}.

% -------------------------------------------------------------------------
% Either list references using the bibliography style file IEEEtran.bst
%\clearpage
%\vspace{-0.1cm}
\bibliographystyle{IEEEtran}
\vspace{-0.1cm}
\begin{spacing}{0.82}
	\vspace{-0.2cm}
	\bibliography{thesis_zhao_waspaa2019}
\end{spacing}

\end{sloppy}
\end{document}